\documentclass[aps,prl,twocolumn,preprintnumbers,superscriptaddress,dblfloatfix,nofootinbib]{revtex4-1}
\pdfoutput=1 
\usepackage[utf8]{inputenc}
\usepackage{lmodern}
\usepackage{amsmath,amssymb}
\usepackage[version=4]{mhchem}
\usepackage{comment}
\usepackage{graphicx}
\usepackage{url}
\usepackage{color}
\usepackage{enumitem}
\usepackage{subfigure}
\usepackage[greek.polutoniko,english]{babel}
\usepackage[dvipsnames]{xcolor}
\usepackage[colorlinks=true,breaklinks=true]{hyperref}
\hypersetup{allcolors=[rgb]{0.0 0.0 0.6},linkcolor=[rgb]{0.75 0.05 0.05}}
\usepackage{bm}
\usepackage{epsfig}
\usepackage{amsmath}
\usepackage{amssymb}
\usepackage{slashed}
\usepackage{color}
\usepackage{accents}
\usepackage{url}
\usepackage[dvipsnames]{xcolor}

\newcommand{\exclude}[1]{}

\begin{document}

\title{Catastrogenesis with unstable ALPs as the origin of the NANOGrav 15 yr
gravitational wave signal}

\author{Graciela B. Gelmini}\email{gelmini@physics.ucla.edu}
\affiliation{Department of Physics and Astronomy, University of California, Los Angeles \\ Los Angeles, California, 90095-1547, USA}

\author{Jonah Hyman}\email{jthyman@g.ucla.edu}
\affiliation{Department of Physics and Astronomy, University of California, Los Angeles \\ Los Angeles, California, 90095-1547, USA}

\date{November 30, 2023}

\begin{abstract}

In post-inflation axion-like particle (ALP) models, a stable domain wall network forms if the model's potential has multiple minima. This system must annihilate before dominating the Universe's energy density, producing ALPs and gravitational waves (a process we dub ``catastrogenesis," or ``creation via annihilation"). We examine the possibility that the gravitational wave background recently reported by NANOGrav is due to catastrogenesis. For the case of ALP decay into two photons, we identify the region of ALP mass and coupling, just outside current limits, compatible with the NANOGrav signal.
\end{abstract}
\maketitle
\setcounter{equation}{0}
\setcounter{figure}{0}
\setcounter{table}{0}
\setcounter{section}{0}
\setcounter{page}{1}
 
\section{Introduction}

The fundamental importance of gravitational waves (GWs) as messengers of the pre-Big Bang Nucleosynthesis (BBN) era, a yet unknown  epoch  of the Universe from which we do not yet have any other remnants, cannot be overestimated. The NANOGrav pulsar timing array collaboration has recently reported the observation of a stochastic gravitational wave 
background~\citep{NANOGrav:2023gor} in 15 years of data, and has examined its possible origin in terms of new physics~\citep{NANOGrav:2023hvm}. They showed that the pre-BBN annihilation of cosmic walls provides a good fit to their signal,
both as the sole source and in combination with a background from a population of inspiraling supermassive black hole binaries (SMBHBs), which is expected to be its primary conventional physics origin~\citep{NANOGrav:2023hvm}. The annihilation produces a peaked spectrum, whose peak frequency $f_{\rm peak}$ is given by the inverse of the cosmic horizon $\simeq t_{\rm ann}$ at annihilation redshifted to the present. In their fit to the wall annihilation model NANOGrav finds~\citep{NANOGrav:2023hvm} a peak frequency
\begin{equation} \label{cf-def}
f_{\rm peak}= c_f 10^{-8}~{\rm Hz}~,
\end{equation}
and a peak energy density 
\begin{equation} \label{cOmega-def}
\left.\Omega_{\rm GW}h^2\right|_{\rm peak}= c_\Omega 10^{-8}~,
\end{equation}
with coefficients $c_f$ and $c_\Omega$ of order 1. In particular $c_\Omega \simeq 1$, while $c_f$ can have larger values. 

Here we consider the annihilation of a $U(1)$ pseudo Nambu-Goldstone boson stable string-wall system as the origin of the NANOGrav signal, based on our previous recent work~\citep{Gelmini:2021yzu, Gelmini:2022nim,Gelmini:2023ngs}, to which we refer often in the following. 

Many extensions of the Standard Model (SM) of elementary particles assume an approximate global $U(1)$ symmetry spontaneously broken at an energy scale $V$. The symmetry is not exact, but explicitly broken at another scale $v \ll V$. Thus the model has a pseudo Nambu-Goldstone boson we denote with $a$,  with mass $m_a\simeq v^2/V$. These models, include the original axion~\citep{Peccei:1977hh, Weinberg:1977ma, Wilczek:1977pj}, invisible axions (also called ``QCD~axions")~\citep{Kim:1979if, Shifman:1979if, Zhitnitsky:1980tq, Dine:1981rt}, majoron models~\citep{Chikashige:1980ui, Gelmini:1980re, Rothstein:1992rh, Gu:2010ys, Lazarides:2018aev, Reig:2019sok, Abe:2020dut, Bansal:2022zpi},  familon models~\citep{Wilczek:1982rv,Reiss:1982sq,Gelmini:1982zz}, and axion-like particles (ALPs) (e.g.~\cite{Svrcek:2006yi,Arvanitaki:2009fg,Acharya:2010zx,Dine:2010cr,Jaeckel:2010ni}). Many models predict a large mass for the QCD axion~\citep{Fukuda:2015ana, Dimopoulos:2016lvn,Agrawal:2017ksf,Gaillard:2018xgk}, including the ``high-quality QCD axion''~\citep{Hook:2019qoh} and previous models (see e.g. Section 6.7 of~\cite{DiLuzio:2020wdo}). Heavy majorons, which could get a mass from soft breaking terms or from gravitational effects (see e.g.~\cite{Rothstein:1992rh,Gu:2010ys,Lazarides:2018aev,Reig:2019sok,Abe:2020dut}), have been considered as well (see e.g.~\cite{Gu:2010ys,Abe:2020dut}), even of mass in the TeV range. Since we need a specific type of model to take into account existing experimental bounds,  we concentrate on ALPs coupled to photons. ALPs are one of the most studied types of dark matter candidates. They are extensively searched for in a variety of laboratory experiments and astrophysical observations, their coupling to photons being one of the most studied as well.

We assume that the spontaneous symmetry breaking happens after inflation, in which case cosmic strings appear during the spontaneous breaking transition, and a system of cosmic walls bounded by strings is produced when the explicit breaking becomes dynamically relevant, when $t \simeq m_a^{-1}$. The cosmic strings then enter into a ``scaling" regime, in which the number of strings per Hubble volume remains of order 1 (see e.g.~\cite{Vilenkin:1984ib} and references therein). The subsequent evolution of the string-wall system depends crucially on the number of minima of the potential after
the explicit breaking, which may be just one minimum, $N=1$, or several, $N > 1$. With $N=1$, ``ribbons" of walls bounded by strings 
surrounded by true vacuum form, which shrink very fast due to the pull of the walls on the strings, leading to the immediate annihilation of the string-wall system (see e.g.~\cite{Gorghetto:2021fsn}).

We concentrate on the $N>1$ case, where the $U(1)$ symmetry is broken into a discrete $Z_N$ symmetry,  in which each string connects to $N$ walls forming a stable string-wall system. A short time after walls form,  when friction of the walls with the surrounding medium is negligible,  the string-wall system enters into another scaling regime in which the linear size of the walls is the cosmic horizon size $\simeq t$. Thus its energy density is $\rho_{\rm walls}\simeq \sigma/t$, where $\sigma$ is the energy density per unit area of the walls. The energy density in this system grows faster with time than the radiation density, and would come to dominate the energy density of the Universe, leading to an unacceptable cosmology~\citep{Zeldovich:1974uw}, unless it annihilates earlier. 

If the  $Z_N$ is also an approximate symmetry, then there is a 
 ``bias," a small energy difference between the $N$ minima, which chooses one of them to be that with minimum energy.  The energy difference between two vacua at both sides of each wall
accelerates each wall toward its adjacent higher-energy vacuum, which drives the domain walls to their annihilation~\citep{Zeldovich:1974uw} (see also e.g.~\cite{Gelmini:1988sf}). 
As in our previous recent work~\citep{Gelmini:2021yzu, Gelmini:2022nim,Gelmini:2023ngs}, we adopt the $Z_N$ explicit breaking term in the scalar potential originally proposed for QCD axions~\citep{Sikivie:1982qv,Chang:1998bq}, and
parameterized as $V_{\rm bias} \simeq \epsilon_b v^4$, with a dimensionless positive coefficient $\epsilon_b \ll 1$.
For small enough $\epsilon_b$ values, ALPs 
are dominantly produced when the string-wall system annihilates, together with GWs, a process that we named ``catastrogenesis"~\citep{Gelmini:2022nim}, after the Greek word \textgreek{katastrof\'h}, for ``overturn'' or ``annihilation." 

The emission of  GWs by the initial system of cosmic strings 
ends when walls appear. Thus,  there is a low-frequency cutoff of 82 $(m_a/{\rm GeV})^{1/2}$ Hz~\citep{Chang:2019mza,Gouttenoire:2019kij,Gorghetto:2021fsn}, corresponding to the inverse of the cosmic horizon when walls appear, redshifted to the present. This is much higher than the relevant frequencies for $m_a\simeq$ GeV, so strings do not contribute to the NANOGrav signal in this model. 

We assume radiation domination during the times of interest. In this case, the present peak GW density is related to the temperature at annihilation $T_{\rm ann}$ by
\begin{equation} \label{f-peak}
f_{\rm peak}\simeq  0.76 \times 10^{-7} \text{Hz} ~\frac{T_{\rm ann}}{\rm GeV}~ \frac{\left[g_\star(T_{\rm ann})\right]^{1/2}}{\left[g_{s \star}(T_{\rm ann}) \right]^{1/3}}~,
\end{equation}
where $g_\star$ and $g_{s \star}$ are the energy and entropy density numbers of degrees of freedom. Thus, Eq.~\eqref{cf-def} also gives
$T_{\rm ann}$ in terms of $c_f$
\begin{equation} \label{Tann-cf}
T_{\rm ann}\simeq 82.5
~ c_f~ {\rm MeV} \left[\frac{16.5}{g_\star(T_{\rm ann})}\right]^{1/2}~\left[\frac{g_{s\star}(T_{\rm ann})}{16.5}\right]^{1/3},
\end{equation}
while in terms of the parameters of our model it is
\begin{equation} \label{Tann}T_{\rm ann} \simeq \frac{2.2 \times 10^9 ~{\rm GeV}}{[g_\star(T_{\rm ann})]^{1/4}}~  \sqrt{\frac{\epsilon_b ~ m_a}{f_\sigma ~ {\rm GeV}}}~.\end{equation}
The peak energy density is
\begin{equation} \label{OmegaGW-walls}
\left.\Omega_{\rm GW}h^2\right|_{\rm peak}\simeq \frac{1.2 \times 10^{-79} \epsilon_{\rm GW}~ g_\star(T_{\rm ann})}{\epsilon_b^{2}~\left[g_{s \star}(T_{\rm ann}) \right]^{4/3}}
\left(\frac{f_\sigma V}{N {\text{GeV}}}\right)^4
\end{equation}
where $f_\sigma$ is a parameter entering into the definition of the energy per unit area of the walls, $\sigma \simeq f_\sigma v^2 V/ N$, and $f_\sigma \simeq 6$ for most assumed potentials. We include  in Eq.~\eqref{OmegaGW-walls} a dimensionless factor $\epsilon_{\rm GW}$ found in numerical simulations (e.g.~\cite{Hiramatsu:2012sc}). When needing to fix its value we use  $\epsilon_{\rm GW}=0.7$ as adopted in the NANOGrav fit~\citep{NANOGrav:2023hvm} following~\cite{Hiramatsu:2013qaa} (in our earlier work we took instead $\epsilon_{\rm GW}=10$, using Fig.~8 of~\cite{Hiramatsu:2012sc}). Since $g_\star= g_{s \star}$ for $T> 1$ MeV, we set them equal in the following.
We address the reader to~\cite{Gelmini:2021yzu, Gelmini:2022nim} for the derivation of these equations. 

Our previous results~\citep{Gelmini:2021yzu, Gelmini:2022nim} show
that the requirement that the ALP density not exceed that of dark matter, $\Omega_a h^2 \lesssim 0.12$, implies
\begin{equation} \label{window}
\frac{\left.\Omega_{\rm GW}h^2\right|_{\rm peak}}{10^{-17}} \left(\frac{f_{\rm peak}}{10^{-9} {\rm Hz}}\right)^2 < 10^{-2}~,
\end{equation} 
so the model cannot produce the NANOGrav signal with stable ALPs.
Thus we concentrate on ALPs that are unstable and decay into SM products that thermalize early enough to leave no trace by the time of Big Bang Nucleosynthesis (BBN), such as we considered in~\cite{Gelmini:2023ngs}. To escape existing laboratory, astrophysical, and cosmological limits on ALP decays into SM products, these ALPs must have a  mass $m_a$ in the GeV range or higher, depending on the decay mode (see e.g.~\cite{Gelmini:2023ngs} and references therein).

Similar or related models have been studied recently in relation to pulsar timing array data, e.g.~\cite{Guo:2023hyp, Madge:2023cak, Kitajima:2023cek, Gouttenoire:2023ftk, Blasi:2023sej, Servant:2023mwt, Ge:2023rce}. ~\cite{Servant:2023mwt} considered the same type of models we study here, but with the purpose of excluding parameter regions disfavored by the NANOGrav 15 yr data, which they analyzed independently. Our purpose is instead to try to explain the signal, and thus we stay away from the disfavored region (shown in gray in the lower left panel of Fig.~12 of~\cite{NANOGrav:2023hvm} and the right panel of Fig.~2 of~\cite{Servant:2023mwt}).

\section{Unstable ALP models that can produce the NANOGrav signal}

In~\cite{Gelmini:2023ngs} we assumed  $m_a$ was sufficiently larger than 1 GeV for ALPs decaying into SM particles to comfortably escape existing experimental limits. However,  we need to be more nuanced here and explore the viability of somewhat lighter ALPs. The reason is that requiring the walls to form at least one order of magnitude in temperature after strings appear, combined with  
upper limits on $T_{\rm ann}$ determined by NANOGrav to explain the signal,  impose $m_a \lesssim 1.8$ GeV, as we are going to show now. 

Walls appear when the Hubble parameter is $H(T_{\rm w}) \simeq m_a/3$, i.e. when the temperature is
\begin{equation} \label{Tw}T_{ \rm w} \simeq \frac{1.6 \times 10^9 ~{\rm GeV}}{\left[g_\star(T_{\rm w})\right]^{1/4}}  \left(\frac{m_a}{\rm GeV}\right)^{1/2}.\end{equation}
Thus $T_{\rm w}$ depends only on $m_a$ ($g_\star(T_{\rm w})\simeq 105$, since $T_{\rm w}> 100$ GeV). As in our previous papers, we consider $m_a$ to be temperature independent. A temperature dependence would not affect the annihilation process, which happens late enough for $m_a$ to have reached its present constant value in any case, but could affect $T_{\rm w}$.

Combining Eqs.~\eqref{cOmega-def} and \eqref{OmegaGW-walls} fixes the ratio $V^2/\epsilon_b$, and Eqs.~\eqref{Tann-cf} and \eqref{Tann} fix the product $\epsilon_b m_a$.  Thus, given Eqs.~\eqref{cf-def} and~\eqref{cOmega-def}, we obtain $V$ as a function of $m_a$,
\begin{equation} \label{V}
V\simeq  \frac{5.0 \times 10^{7}~\rm{GeV}}{\epsilon_{\rm GW}^{1/4}} \frac{N}{f_\sigma^{1/2}} \left(\frac{\rm GeV}{m_a}\right)^{1/2} c_f c_\Omega^{1/4} \left[\frac{g_\star(T_{\rm ann})}{16.5}\right]^{1/6},
\end{equation}
and consequently $m_a$ in terms of the ratio $T_{\rm w}/V$, 
\begin{equation} \label{ma-upper-function}
 m_a \simeq \frac{(T_{\rm w}/V)}{0.1} c_f c_\Omega^{1/4} \frac{N}{f_\sigma^{1/2}} \frac{10.3
 ~ {\rm MeV}}{\epsilon_{\rm GW}^{1/4}}\left[\frac{g_\star(T_{\rm ann})}{16.5}\right]^{1/6}. 
\end{equation}
We require $T_{\rm w} / V \lesssim 0.1$   so walls form  at least one order of magnitude in temperature after strings appear. Larger values of $N$ are favorable to allow larger $m_a$. To our knowledge, upper limits on $N$  have been studied only in QCD axion models ~\citep{DiLuzio:2017pfr}  in which $N= 20$ is possible. For axions coupled to gluons, $N$ is given by the color anomaly coefficient. Similarly, non-perturbative effects in a dark sector~\citep{Arvanitaki:2009fg, Jaeckel:2010ni}, generically lead to $N>1$ for ALPs, thus possibly to similarly large $N$ values.  We will thus adopt $N= 20$. Replacing also $f_\sigma=6$ and $\epsilon_{\rm GW}=0.7$,
\begin{equation} \label{ma-upper}
 m_a \simeq \frac{(T_{\rm w}/V)}{0.1} c_f c_\Omega^{1/4}  ~92~ {\rm MeV} \left[\frac{g_\star(T_{\rm ann})}{16.5}\right]^{1/6}.  
\end{equation}
An upper limit on $c_f$ thus provides an upper limit on $m_a$ and vice versa (since the NANOGrav fit prefers  $c_\Omega \simeq$ 1~\citep{NANOGrav:2023hvm}). Looking in Fig.~12 of~\cite{NANOGrav:2023hvm}, the range of annihilation temperatures (called $T_\star$ in that paper) where the NANOGrav signal can be explained by the annihilation of domain walls into SM products (DW-SM, the model most similar to ours), we can see that $T_{\rm ann} \lesssim$ 1 GeV (close to the upper boundary of the red region in the figure). By Eq.~~\eqref{Tann-cf}, this corresponds to $c_f \lesssim 15$ (taking into account the rapid change of $g_\star$ values for temperatures in the 100 MeV range, $g_\star (1 ~ \rm{GeV}) \simeq 70$). Through Eq.~\eqref{ma-upper}, this implies $m_a \lesssim 1.8$ GeV.
A more conservative upper limit on the annihilation temperature is the upper boundary of the 95\% credible interval including a SMBHB contribution quoted in the text of~\cite{NANOGrav:2023hvm},  843 MeV. This implies through Eq.~\eqref{Tann-cf} (with $g_\star(0.84~{\rm GeV}) \simeq 
68$)  $c_f \lesssim 13$, and through Eq.~\eqref{ma-upper} $m_a \lesssim 1.5 ~ \rm{GeV}$.

Let us now consider the experimental limits on ALPs coupled to photons through a Lagrangian term
\begin{equation}
\mathcal{L}_{a \gamma \gamma} = \frac{c_{\gamma \gamma}}{f_a} a F_{\mu\nu} \tilde{F}^{\mu\nu}~,
\end{equation}
where $F_{\mu\nu}$ is the electromagnetic field tensor and $\tilde{F}^{\mu\nu}$ its dual,
$|c_{\gamma\gamma}|$ is a dimensionless coupling constant, and $f_a= V/N$ is given by Eq.~\eqref{V} divided by $N$, and is independent of $N$, thus
\begin{equation} \label{inverse-fa}
\frac{1}{f_a}\simeq \frac{1.4 \times 10^{-8}}{c_f c_\Omega^{1/4} {\rm GeV}} \left(\frac{m_a}{100 ~{\rm MeV}}\right)^{1/2} \left[\frac{16.5}{g_\star(T_{\rm ann})}\right]^{1/6}.
\end{equation}
Or using  Eq.~\eqref{ma-upper} in Eq.~\eqref{inverse-fa}, 
\begin{equation} \label{inverse-fa-cf}
\frac{1}{f_a}  
\lesssim  
\frac{1.4 
\times 10^{-8}}{(c_f c_\Omega^{1/4} )^{1/2}~ \rm{ GeV}}  \left[\frac{16.5}{g_\star(T_{\rm ann})}\right]^{1/12}.
\end{equation}
So a larger 
 $c_f$ (thus also a larger $T_{\rm ann}$) makes the coupling smaller.  

Requiring the upper limit on the ALP mass in Eq.~\eqref{ma-upper} to reach $m_a \simeq$ 300 MeV (to avoid experimental limits on lighter ALPs shown in Fig.~1), we obtain $c_f \gtrsim 2.9$ and $T_{\rm ann} \gtrsim$ 200 MeV (thus $g_\star \simeq 42$), which implies $1/f_a< 7.5 \times 10^{-9}/ {\rm GeV}$.
Requiring instead the upper limit to be $m_a \simeq 1.8$ GeV, which corresponds to $T_{\rm ann} \simeq$ 1 GeV since as mentioned above, $c_f= 15$, with  $g_\star(1~ {\rm GeV}) \simeq 70$, Eq.~\eqref{inverse-fa} or~\eqref{inverse-fa-cf} implies $1/f_a< 3.1 \times 10^{-9}/ {\rm GeV}$. 

Assuming $|c_{\gamma\gamma}| \lesssim $ 1, these upper limits on $1/f_a$ translate into upper limits on the ALP coupling to photons as a function of $m_a$. These limits constitute the upper boundary of the gray region in Fig.~1. Fig.~1 also shows relevant regions rejected by the most up-to-date limits on ALPs. Notice that if $|c_{\gamma\gamma}| > 1$, the region extends upward (as indicated by the dashed lines) where experimental limits (not only astrophysical limits) become important.

The value of $|c_{\gamma\gamma}|$ depends on the completion of the ALP model. It has been extensively studied only for the QCD axion, where $|c_{\gamma\gamma}| \simeq \alpha_{\rm EM}/8 \pi \simeq 2.9 \times 10^{-4}$ in the simplest models. However, $|c_{\gamma\gamma}|$ can be many orders of magnitude, even exponentially, larger in some models (see e.g. the ``Axions and Other Similar Particles" review in~\cite{ParticleDataGroup:2022pth} or~\cite{Farina:2016tgd, Agrawal:2017cmd, Daido:2018dmu,DiLuzio:2020wdo,Plakkot:2021xyx}). Notice that with the $|c_{\gamma\gamma}|$ value in the simplest QCD axion models, the upper boundary of our region of compatibility (gray) in Fig.~1 would move to the dot-dashed line in the region excluded by BBN limits (yellow), i.e. the region of compatibility would not exist. 

 \begin{figure}[t]\centering\vspace{0em}\hspace{-1em}
\includegraphics[width=0.49\textwidth]{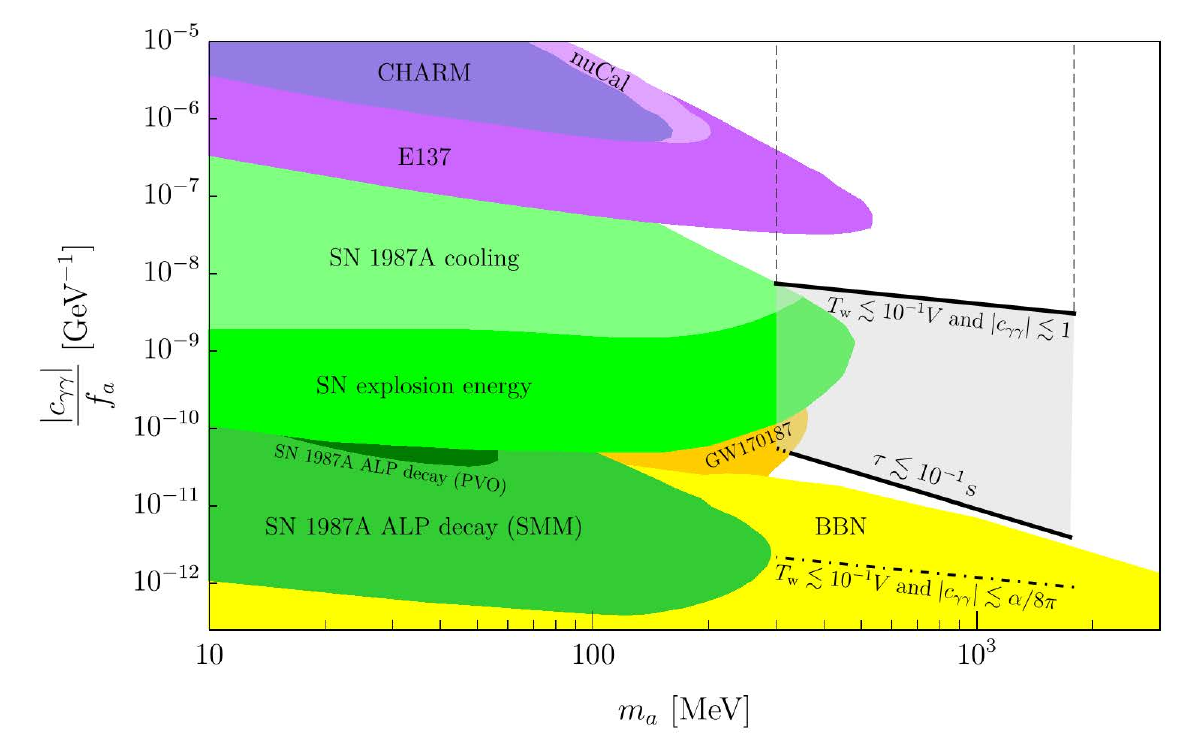}
\caption{Region (in gray) of ALP coupling to two photons versus ALP mass $m_a$ for models which could explain the NANOGrav 15 yr signal, where  300 MeV $<m_a<$ 1.8 GeV, together with (colored) relevant regions excluded by: SN 1987A cooling~\protect\citep{Caputo:2022mah, Caputo_2022b}, SN 1987A ALP decay (Solar Maximum Mission)~\protect\citep{müller2023investigating}, SN 1987A ALP decay (Pioneer Venus Orbiter)~\protect\citep{Diamond_2023}, supernovae (SN) explosion energy~\protect\citep{Caputo:2022mah}, GW170817~\protect\citep{diamond2023multimessenger}, BBN + $N_{\rm{eff}}$ limits~\protect\citep{Depta_2020}, and experimental limits~\protect\citep{Dolan:2017osp,Dobrich:2019dxc}. This figure reproduces a portion of Fig.~9 of~\protect\cite{Antel:2023hkf} with additions from~\protect\cite{AxionLimits}.
}
\label{fig:differentialspectrum}
\end{figure}

We will now check the lifetime and the fraction of the density constituted by ALPs at the time of decay. The decay rate (see e.g. Eq.~(138) of~\cite{Antel:2023hkf})
\begin{equation} 
\label{decay-rate-eq}
\Gamma (a \to \gamma \gamma) =  \frac{|c_{\gamma\gamma}|^2 m_a^3}{4 \pi f_a^2} 
\end{equation}
corresponds to a lifetime (using Eq.~\eqref{inverse-fa})
\begin{equation} \label{lifetime}
\tau=  
\frac{c_f^2 c_\Omega^{1/2}}{|c_{\gamma\gamma}|^2}4.2 \times 10^{-5} {\rm ~sec} \left(\frac{100 ~{\rm MeV}}{m_a}\right)^4 \left[\frac{g_\star(T_{\rm ann})}{16.5}\right]^{1/3}~. 
\end{equation}
With $|c_{\gamma\gamma}|$ in the range 0.1 to 1, we can have $\tau \simeq t_{\rm ann}$, i.e. the decay can happen at annihilation. Requiring $\tau \lesssim 0.1$ sec, so that the decay happens early enough not to affect BBN, translates through Eq.~\eqref{decay-rate-eq} into $|c_{\gamma\gamma}|/f_a \gtrsim 0.9 \times 10^{-11} ({\rm GeV}/ m_a)^{3/2}/ {\rm GeV}$.
This determines the lower boundary of the gray region shown in Fig.~1 (where $|c_{\gamma\gamma}|$ goes from $\simeq 10^{-3}$ for $m_a\simeq 1.8$~GeV to $\simeq 10^{-2}$ for $m_a \simeq 0.3$~GeV).

To compute the density of the string-wall system with respect to that of radiation at annihilation, we consider that, had the system not annihilated, its energy density $\rho_{\rm{walls}} \simeq \sigma/t $ would have continued to grow until the moment we call wall-domination $t_{\rm{wd}}$, at which it becomes as large as the radiation energy,  $\rho_{\rm{walls}}(t_{\rm{wd}}) \simeq \rho_{\rm{rad}}(t_{\rm{wd}})$. The temperature of wall-domination is 
(see~\cite{ Gelmini:2022nim,Gelmini:2023ngs})
\begin{equation} \label{T wall domination-2} T_{\rm wd} \simeq \frac{3.4~ {\rm GeV}}{[g_\star(T_{\rm wd})]^{1/4}}\frac{f_\sigma^{1/2}}{N}\left(\frac{V}{\rm 10^9~GeV}\right)\left(\frac{m_a}{\rm 10~GeV}\right)^{1/2}.\end{equation}
Besides,
$\rho_{
\rm{walls}}(t_{\rm ann})/ \rho_{\rm{walls}}(t_{\rm{wd}}) \simeq t_{\rm{wd}}/ t_{\rm {ann}}$, and the 
ratio of radiation densities at wall-domination and annihilation is given by
the ratio of $g_\star T^4$ at each temperature.  Combining these equations we find
\begin{equation}
\label{density-ratio-annih}
    \frac{\rho_{\rm{walls}}(T_{\rm ann})}{\rho_{\rm{rad}}(T_{\rm ann})} \simeq 
    \left(\frac{g_\star(T_{\rm{wd}})}{g_\star(T_{\rm ann})}\right)^{1/2}
    \left(\frac{T_{\rm{wd}}}{T_{\rm ann}}\right)^2~.
 \end{equation}
Using Eq.~(\ref{Tann-cf}) and Eq.~(\ref{V}) in Eq.~(\ref{T wall domination-2}), we find
\begin{equation}\label{density-ratio-annih-value}\frac{\rho_{\rm{walls}}(T_{\rm ann})}{\rho_{\rm{rad}}(T_{\rm ann})} \simeq 0.13 ~ c_\Omega^{1/2} \left(\frac{g_\star(T_{\rm ann})}{16.5}\right)^{1/6},\end{equation}
which shows that this ratio is always $<1$ for the annihilation temperatures we consider. Since practically all the density in the string-wall system goes into nonrelativistic (or quasi-nonrelativistic) ALPs at annihilation, considering the redshift of the ALP and radiation densities until ALPs decay at temperature
$T_{\rm decay}$, 
\begin{equation}
\label{density-ratio-decay}
    \frac{\rho_{\rm ALPs}(T_{\rm decay})}{\rho_{\rm rad}(T_{\rm decay})} \simeq 
     \left(\frac{T_{\rm ann}}{T_{\rm decay}}\right)\frac{\rho_{\rm walls}(T_{\rm ann})}{\rho_{\rm rad}(T_{\rm ann})}.
 \end{equation}
As we mentioned above, $T_{\rm decay}$ can be very close to $T_{\rm ann}$, so ALPs do not get to matter dominate in our model and the decays happen early enough for the products to thermalize long before BBN. Otherwise, there would be a period of ALP matter domination before ALPs decay, which is in principle not problematic since the decays happen much before BBN, but would be a scenario deserving further study. 

The range of $c_f$ values, 2.9 to 15, that we have found above corresponds to a peak frequency range through Eq.~\eqref{cf-def}.  In Fig.~2 we indicate two approximate spectra, with the maximum and minimum  $f_{\rm peak}$ in the mentioned range. Frequencies $f < f_{\rm peak}$ correspond to
super-horizon wavelengths at annihilation, so causality requires a $\sim f^3$ dependence~\citep{Caprini:2009fx} for wavelengths that enter into the horizon during radiation domination, see e.g.~\cite{Barenboim:2016mjm,Cai:2019cdl, Hook:2020phx}.
For $f > f_{\rm peak}$ the spectrum depends instead on the particular production model.~\cite{Hiramatsu:2012sc} finds a  roughly $1/f$ dependence (although the approximate slope slightly depends on $N$), which we use for Fig.~2. In Fig.~2 the rough signal region of NANOGrav, as well as the limits and reach of other GW observatories, is shown. 

 \begin{figure}[t]\centering\vspace{0em}\hspace{-1em}
\includegraphics[width=0.49\textwidth]{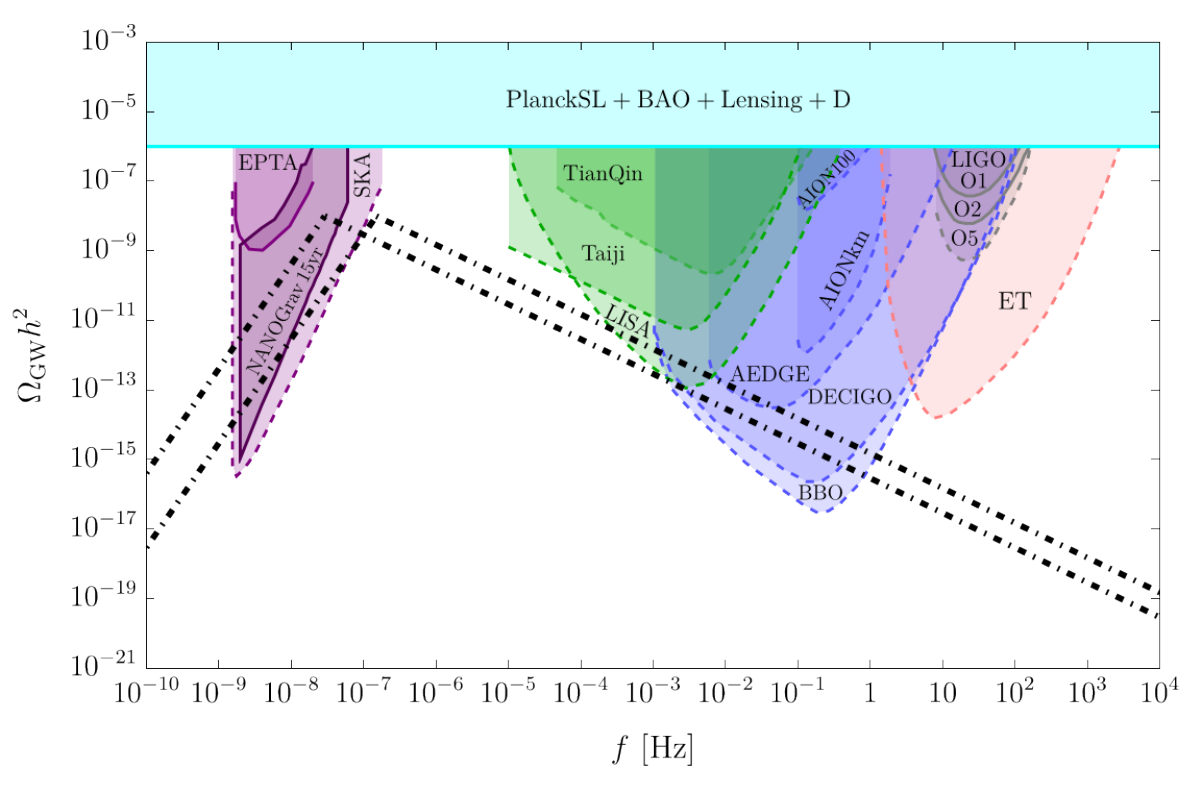}
\caption{Approximate spectra which could account for the NANOGrav signal in our catastrogenesis model, with peak amplitude $\left. \Omega_{\rm{GW}}h^2 \right|_{\rm{peak}}= 10^{-8}$ and peak frequencies $f_{\rm peak} = 2.9 \times 10^{-8}$~Hz and  $1.5 \times 10^{-7}$~Hz,  which are the minimum and maximum frequencies we found (see text). Allowed spectra would have peak frequencies in between these two. Also shown are the approximate NANOGrav 15 yr signal~\protect\citep{NANOGrav:2023hvm} (in purple) and limits (solid line boundaries) or reach (dashed line boundaries) of other GW detectors: the European Pulsar Timing Array (EPTA)~\protect\citep{Antoniadis:2023rey} and the Square Kilometre Array (SKA)~\protect\citep{Janssen:2014dka} in purple; the space-based experiments TianQin~\protect\citep{TianQin:2015yph}, Taiji~\protect\citep{Ruan:2018tsw}, and the Laser
Interferometer Space Antenna (LISA)~\protect\citep{LISA:2017pwj} in green; the Atom Interferometer
Observatory and Network (AION)~\protect\citep{Badurina:2019hst}, the Atomic Experiment for Dark Matter and Gravity Exploration in Space (AEDGE)~\protect\citep{AEDGE:2019nxb}, the Deci-hertz Interferometer Gravitational wave
Observatory (DECIGO)~\protect\citep{Seto:2001qf}, and the Big Bang Observer (BBO)~\protect\citep{Corbin:2005ny} in blue; the ground-based Einstein Telescope (ET) in red~\protect\citep{Sathyaprakash:2012jk}; and the Laser Interferometer Gravitational-wave Observatory (LIGO) in gray~\protect\citep{LIGOScientific:2019vic}. The cyan band corresponds to the 95\% C.L. upper limit on the effective
number of degrees of freedom during CMB emission from Planck and other data~\protect\citep{Pagano:2015hma}, which imposes $\Omega_{\rm GW} h^2 < 10^{-6}$.
}
\end{figure}

\section{Possibility of primordial black hole formation}

The formation of primordial black holes (PBHs) during the process of annihilation of the string-wall system is an exciting possible aspect of ALP models with $N>1$. We recently dealt with the possibility of producing ``asteroid-mass" PBHs, in the range in which they could constitute all of the dark matter, in~\cite{Gelmini:2023ngs}.  If formed, the PBH mass in the  models in the present paper would be in the range of $0.1$ to a few solar masses, but PBH abundance would be too large to be allowed, and this would reject these models.

However, the formation of PBHs is uncertain. The argument for PBH formation, first presented in~\cite{Ferrer:2018uiu} for QCD axions, is that in the latest stages of wall annihilation in $N>1$ models ($t> t_{\rm ann}$) closed walls could arise and collapse in an approximately spherically symmetric way. In this case, if the characteristic  linear size of the walls continues to grow with time after annihilation starts,  some fraction of the closed walls could reach their Schwarzschild radius $R_{\rm Sch}$ and collapse into PBHs. The figure of merit used is $p(t)= R_{\rm Sch}/t ={2 M(t)}/({t\, M_{\rm P}^2})$, where $M_{\rm P}$ is the Planck mass and $M(t)$ is the mass within the collapsing closed wall at time $t$. Reaching $p(t)=1$  would indicate the formation of PBHs. This definition is based on the fact  that while walls are in the scaling regime, the linear size of the walls $L$ is close to the horizon size ($L \simeq t$).

Annihilation starts when surface tension of the walls, which produces a pressure $p_T \simeq \sigma/t$ that decreases with time (which tends to rapidly straighten out curved walls to the horizon scale $t$), is compensated by the volume pressure $p_V \simeq V_{\rm bias}$ (which tends instead to accelerate the walls toward their higher-energy adjacent vacuum). In our model, $p_V \ll  p_T$ when walls form. At a later time, when $p_T \simeq p_V$, the bias drives the walls (and the strings bounding them) to annihilate within a Hubble time. This defines $t_{\rm ann} \simeq \sigma/V_{\rm bias}$, after which  $V_{\rm bias}$ dominates the energy density. 
After annihilation starts, $L \simeq t$ is no longer guaranteed.

We have checked that for the models in this paper, we always have $p(t_{\rm ann}) < 1$. If $L$ continues being close to $t$ for $t > t_{\rm ann}$, then $p(t > t_{\rm ann})\simeq V_{\rm bias} L^3/ L$ grows with time as $t^2$ and eventually reaches 1. However,  if $L$ decreases with time at some point after annihilation starts, the figure of merit may never reach 1. Based on the simple power-law parameterization we used in our previous recent work~\citep{Gelmini:2022nim,Gelmini:2023ngs} for the evolution of the energy density after annihilation starts, 
namely ${\rho_{\rm walls}(T)}/{\rho_{\rm walls}(T_{\rm ann})} \simeq \left({T}/{T_{\rm ann}}\right)^\alpha$ (with a parameter $\alpha$ that needs to be extracted from simulations), we can make a naive estimate of how the characteristic linear wall size $L$ within a Hubble volume $t^3$ evolves with time. In Appendix A we show how this naive estimate requires $\alpha <6$ for $L$ to ever become larger than $t_{\rm ann}$. 
The only simulations of the annihilation process available~\citep{Kawasaki:2014sqa}  find $\alpha \geq 7$~\citep{Gelmini:2022nim,Gelmini:2023ngs}. On the other hand, they also seem to indicate that the evolution of the string-wall system continues being close to that in the scaling regime for some time. Therefore, more detailed simulations of the annihilation process are required to elucidate the appearance of PBHs.

In addition, a large enough departure from spherical symmetry due to angular momentum or vacua with different energy on different sides of the collapsing closed wall could prevent the formation of PBHs. Since the formation of PBHs is such an uncertain consequence of ALP models with $N>1$, we do not use this feature to reject any of these models.

\section{Conclusions}

We pointed out that the recently confirmed stochastic gravitational wave background could be due to pseudo Nambu-Goldstone bosons, whose existence could only be revealed through their decays and this background. In particular, we examined unstable ALP models which can produce the recent NANOGrav 15 yr signal. ALP models have a complex cosmology in which a stable system of walls bounded by strings develops (for $N>1$), and nonrelativistic ALPs and gravitational waves are produced when the cosmic string-wall system annihilates (a process we dubbed ``catastrogenesis" in our recent work on these models). The annihilation produces a distinctive peaked spectrum, at a frequency corresponding to the inverse of the cosmic horizon at annihilation. Thus, this peak frequency is related to the annihilation temperature.

We require ALPs to decay into Standard Model (SM) products which thermalize much before BBN. In particular, we have shown that ALPs decaying into two photons in the region of masses and couplings necessary to explain the signal can evade existing observational limits, the most relevant of which are derived from supernova data (see Fig.~1), for  ALP masses from about 300 MeV to 1.8 GeV.  The model closest to ours that  NANOGrav fitted to their signal is that of domain walls decaying into SM products (DW-SM).  Our model is very similar to this one if the ALP decay happens very shortly after string-wall system annihilation, which we showed is possible. Thus we use the NANOGrav fits to this model to select a range of annihilation temperatures and thus peak frequencies.    

We have found a range of $c_f$ values (as defined in Eq.~\eqref{cf-def})
which corresponds to the range of peak frequencies  from $f_{\rm peak}= 2.9 \times  10^{-8}$ Hz to $f_{\rm peak}= 1.5 \times  10^{-7}$ Hz.
This corresponds to annihilation temperatures in the range 200 MeV to 1 GeV. This temperature range overlaps with the upper portion of the 68\% credible interval (which goes to 275 MeV) and the 95\% credible interval (which goes to 505 MeV) quoted by NANOGrav~\citep{NANOGrav:2023hvm} if their DW-SM model is the sole origin of the signal. Considering their fit done with the addition of a SMBHB contribution, our temperature range overlaps with a larger portion of both the 68\% credible interval (which goes to 309 MeV) and the 95\% credible interval (which goes to 843 MeV) quoted in the text, and is included within the red region in the lower left corner of Fig.~12 of~\cite{NANOGrav:2023hvm} (for its DW-SM + SMBHB fit).

The upper portion of the region of ALP-photon coupling and mass necessary to explain the NANOGrav signal (shown in Fig.~1), for couplings above $10^{-7}$ GeV$^{-1}$, could be tested in the future by DarkQuest, HIKE-dump and SHiP, as shown e.g. in Fig.~133 of~\cite{Antel:2023hkf} (see references therein). The lower portion would be tested if a new supernova is observed by the Supernova Early Warning System (SNEWS)~\citep{SNEWS:2020tbu}, which would allow to extend considerably all the limits derived from SN1987A.

\section*{Acknowledgements}
We thank E. Vitagliano for useful comments. The work of GG was supported in part by the U.S. Department of Energy (DOE) Grant No. DE-SC0009937.

\appendix

\section{Appendix A}

 Before annihilation starts, the energy density of the walls in the scaling regime is $\rho_{\rm walls} \simeq \sigma/t \gg V_{\rm bias}$. The annihilation of the string-wall system starts when the bias volume energy density, or magnitude of volume pressure, $V_{\rm bias}$ becomes of the same order as  the energy density, or surface tension, of the walls $\sigma/t$ ($t_{\rm ann} \simeq \sigma/V_{\rm bias}$), after which $V_{\rm bias}$ dominates and accelerates walls towards the higher-energy vacuum adjacent to each wall. If PBHs do not form at annihilation, i.e. $p(t_{\rm ann}) < 1$, the energy contained in a closed wall will need to increase with time for PBHs to form later, at a time $t_\star$ such that $p(t_\star) =1$. Since the energy density $V_{\rm bias}$ is constant, this requires that the dimensions of the closed walls keep growing for $t> t_{\rm ann}$. In fact, if the characteristic linear dimension $L$ of walls continues being close to $t$, $L\simeq t$, then $p(t > t_{\rm ann})\sim V_{\rm bias} L^3/ L$ grows with time as $t^2$ and eventually reaches 1. However,  if $L$ decreases with time and never becomes larger than $t_{\rm ann}$, the figure of merit $p(t)$ decreases after annihilation starts and never reaches 1.

Based on the simple power-law parameterization we used in our previous recent work~\citep{Gelmini:2022nim,Gelmini:2023ngs} for the evolution of the energy density after annihilation starts, for $T < T_{\rm ann}$,
\begin{equation}
 \frac{\rho_{\rm walls}(T)}{\rho_{\rm walls}(T_{\rm ann})} \simeq \left(\frac{T}{T_{\rm ann}}\right)^\alpha \simeq \left(\frac{t_{\rm ann}}{t}\right)^{\alpha/2} , 
\end{equation}
with a real positive power $\alpha$ that needs to be extracted from simulations of the annihilation process,
we can make a naive estimate of how the characteristic linear wall size $L$ within a Hubble volume $t^3$ evolves with time. It is easy to do it either assuming  that walls dominate the energy density, or that volume density dominates. In both cases we find the same condition on $\alpha$ for $L$ to continue growing with time, i.e. $L> t_{\rm ann}$ for $t>t_{\rm ann}$. Therefore it is reasonable to assume that the same condition holds in the transition period, when both volume and walls contribute significantly to the energy density of the annihilating string-wall system. 

If the energy in walls still dominates 
\begin{equation}
  \left(\frac{t_{\rm ann}}{t}\right)^{\alpha/2} \simeq \frac{\rho_{\rm walls}(T)}{\rho_{\rm walls}(T_{\rm ann})}\simeq \frac{\sigma L^2}{t^3} \frac{t_{\rm ann}^3}{\sigma t_{\rm ann}^2}\simeq \frac{L^2}{t^3} t_{\rm ann}  . 
\end{equation}
Thus
\begin{equation}
 L= t \left(\frac{t_{\rm ann}}{t}\right)^{(\alpha-2)/4} \end{equation}
and requiring $L/t_{\rm ann} >1$ for $t>t_{\rm ann}$, means that $(t/t_{\rm ann})^{(6-\alpha)/4} >1$, i.e. ${(6-\alpha)/4}>0$, thus $\alpha<6$.

A similar calculation can be done assuming volume energy dominates, $\rho_{\rm walls} (t) \simeq V_{\rm bias} L^3/ t^3$ to find
\begin{equation}
 L= t \left(\frac{t_{\rm ann}}{t}\right)^{\alpha/6} \end{equation}
and requiring $L/t_{\rm ann} >1$ for $t>t_{\rm ann}$, means that $(t/t_{\rm ann})^{(6-\alpha)/6} >$ 1,  with ${(6-\alpha)/6}>0$,  i.e. $\alpha<6$ again. 

In both cases we find the condition $\alpha<6$ for $L$ to become larger than $t_{\rm ann}$ after annihilation stars. However the only simulations available to estimate values of $\alpha$~\citep{Kawasaki:2014sqa} lead to $\alpha \geq 7$ (see~\cite{Gelmini:2022nim,Gelmini:2023ngs} for details). In this case, with our naive estimates the linear size of walls would decrease with time after annihilation starts and PBHs would not form if $p(t_{\rm ann}) < 1$. 

\bibliographystyle{bibi}
\bibliography{bibliography}
 
\end{document}